\documentclass[aps,prc,reprint,floatfix]{revtex4-1}

\usepackage{graphicx}
\usepackage{xcolor}
\usepackage[colorlinks=true, urlcolor=blue, citecolor=blue, linkcolor=blue]{hyperref}
\newcommand{\etal}{\textit{et al}.\ }

\begin{document}

\title{\sf{This is an author-created version of an article published in Physical Review B {\bf 98}, 104508  (2018).\\ The Version of Record is available online at \href{https://doi.org/10.1103/PhysRevB.98.104508}{https://doi.org/10.1103/PhysRevB.98.104508}.\\
\textcopyright 2018 American Physical Society}
\vskip 2pt
\hrule
\vskip 2pt
\colorbox{gray}{\textcolor{white}{\sf \small{Editor's Suggestion}}}
\vskip 20pt {\bf Transverse vortex commensurability effect and sign change of the Hall voltage in superconducting YBa$_{2}$Cu$_{3}$O$_{7-\delta}$ thin films with a nano-scale periodic pinning landscape}}

\author{G.~Zechner}

\author{W.~Lang}
\email[Corresponding author: ]{wolfgang.lang@univie.ac.at}

\affiliation{University of Vienna, Faculty of Physics, Electronic Properties of Materials, Boltzmanngasse 5, A-1090, Wien, Austria}

\author{M.~Dosmailov}
\altaffiliation[Present address: ]{Al-Farabi Kazakh National University, Almaty, Kazakhstan}
\author{M.~A.~Bodea}

\author{J.~D.~Pedarnig}

\affiliation{Johannes-Kepler-University Linz, Institute of Applied Physics, Altenbergerstrasse 69, A-4040 Linz, Austria}

\begin{abstract}
{\footnotesize (Received 10 July 2018; revised manuscript received 3 September 2018; published 18 September 2018)}
\vskip 15pt
The transverse (Hall) voltage in thin films of the high-temperature superconductor YBa$_{2}$Cu$_{3}$O$_{7-\delta}$ with an artificial periodic pinning array is investigated. Columnar defect regions along the crystallographic $c$ axis, in which superconductivity is suppressed, are created by irradiation with He$^+$ ions through a silicon stencil mask. The commensurate arrangement of magnetic flux quanta with the artificial defect lattice is confirmed by maxima of the critical current and minima of the resistance, respectively. The magnetic field dependence of the transverse voltage reveals a commensurability effect characterized by a narrow peak of the Hall coefficient with reversed polarity compared to the background signal. This signature of vortex matching disappears at larger vortex velocities substantiating its close connection with enhanced pinning of vortices at the periodic pinning landscape.\\
\\
DOI: \href{https://doi.org/10.1103/PhysRevB.98.104508}{10.1103/PhysRevB.98.104508}

\end{abstract}

\maketitle

\section{Introduction}

In a current-carrying type-II superconductor in the mixed state vortices can be accelerated by the Lorentz force. The vortices then move in a direction perpendicular to current and magnetic field, giving rise to a voltage drop along the current direction. It causes dissipation in the material, which, despite its different origin, has some similarities with an ohmic resistance. If the vortex trajectories are deflected from their perpendicular-to-current orientation, the resulting electric field gains a component transversal to the current and this leads to a transverse voltage with similar features than the normal-state Hall effect. For that reason it is termed ``vortex Hall effect.''

The close connection between normal-state properties and dissipation due to vortex motion is explained by the models of Bardeen and Stephen \cite{BARD65} and Nozi\`eres and Vinen \cite{NOZI66}, which indeed find that the transport properties of the vortices' normal-state cores and the density of vortices determine the observable longitudinal and transverse voltages. Possible complications become already evident by the fact that the two above-mentioned theories do not fully agree on the vortex Hall effect and, even more, in a real system the vortex dynamics are influenced by additional forces  \cite{KOPN02R} with the Magnus force as one of the prominent examples \cite{SONI97,AO93,ZHU97}. But not only is the equation of motion of a single vortex a source of still ongoing discussion, the importance of vortex many-body effects has been pointed out, too \cite{AO98}.

The discovery \cite{GALF88} that the vortex Hall effect can exhibit an opposite polarity than the hole like normal-state Hall effect in underdoped and optimally doped \cite{NAGA98}  copper-oxide high-$T_c$ superconductors (HTSCs) is in striking contrast to the  traditional models for vortex dynamics \cite{BARD65,NOZI66,KOPN02R,SONI97}. In a temperature down sweep, a sign reversal of the Hall coefficient $R_H$ below the critical temperature $T_c$ appears with precursor effects already visible above $T_c$ in the superconducting fluctuation range \cite{LANG94}. Several theoretical models have attempted to explain this puzzling observation and, based on a renormalized Ginzburg-Landau model for superconducting order parameter fluctuations \cite{NISH97}, the experimental observations could be satisfyingly modeled \cite{PUIC04}. In these theories, subtleties of the Fermi surface determine the sign of the vortex Hall effect \cite{FUKU71,DORS92b,OTTE95,BERG15}.

In a different approach, vortex pinning as the origin of the reversed polarity of the vortex Hall effect has been discussed \cite{WANG91,VINO93,WANG94b,KOPN99,NAKA11}. Also, the dimensionality of the pinning centers can influence the sign of the vortex Hall effect, whether the system is near a Bose or vortex glass transition, respectively \cite{IKED99}. Indeed, the intrinsic strong pinning in near-optimally doped thin films of YBa$_2$Cu$_3$O$_{7-\delta}$ (YBCO) leads to an additional sign reversal---back to positive values as in the normal state---that emerges in low magnetic fields $B < 100$~mT only, when the vortex density is low enough to enable efficient pinning on twin boundaries. \cite{GOB00} Note that this effect is readily canceled by depinning of vortices in high current densities \cite{LANG01} or by tilting the magnetic field off the crystallographic $c$ axis \cite{DANN98,GOB00} and, thus, off the twin-boundary direction, rendering them less efficient for pinning. Nevertheless, a conclusive agreement on the microscopic origin of the vortex Hall effect has not yet been achieved, in particular not for its sign reversal in HTSCs.

The emergence of advanced nanopatterning methods has revived the interest in vortex dynamics due to the intriguing possibility to probe it in an artificially created regular pinning landscape \cite{MOSH10M}. If the vortices are immobile, commensurate arrangements with respect to the defect lattice have been demonstrated with Lorentz microscopy in a superconducting Nb film \cite{HARA96b}. Such ``vortex matching'' effects also appear in magnetization \cite{BAER95} and critical current measurements \cite{LYKO93}, but can be also found in the dynamic case of resistivity measurements that exhibit commensurability minima \cite{FIOR78}.

The interplay between pinned and mobile vortices has some parallels to the insulator-to-metal transition of charge carriers and has been interpreted in terms of a vortex Mott insulator-to-metal transition \cite{NELS93}. Experimental support for this concept has been reported in superconductors with regular pinning arrays by magnetic measurements \cite{BAER95,GOLD09} and via resistivity measurements \cite{JIAN04,POCC15}.

In this paper, we want to explore commensurability effects in the Hall signal. Since the transverse voltage represents a non dissipative contribution to vortex motion it would represent a fundamentally different manifestation of commensurability effects. We shall demonstrate that the Hall signal in YBCO films with a periodic pin array shows a remarkable peak at the matching field, which is comparable to the one in the critical current and is accompanied by a sign change of the Hall voltage.

\section{Experimental techniques}

Thin films of YBa$_2$Cu$_3$O$_{7-\delta}$ are grown epitaxially on (100) MgO single-crystal substrates by pulsed-laser deposition using 248-nm KrF-excimer-laser radiation at a fluence of 3.2~J/cm$^2$. The thickness of the films is $t_z = (210 \pm 10)$~nm as determined by atomic force microscopy. The electrical transport measurements are performed on lithographically pre-patterned bridges with dimensions $240 \times 60\ \mu \mathrm{m}^2$. Two pairs of contacts, allowing for simultaneous acquisition of longitudinal and transverse voltages, are applied on side arms of the bridges using sputtered Au pads. The distance of the longitudinal voltage probes is $100\ \mu \mathrm{m}$. The as-prepared samples had critical temperatures $T_c \sim 90$~K,  transition widths $\Delta T_c \sim 1$~K, and critical current densities from $j_c \sim 3\ \mathrm{to}\  4$~MA/cm$^2$ at 77~K in self-field.

The artificial pinning landscapes consist of columnar defects (CDs) that are created by irradiation with 75~keV He$^+$ ions at a fluence of $3 \times 10^{15}\; \mathrm{cm}^{-2}$. Employing a masked ion beam structuring (MIBS) technique, described in more detail elsewhere \cite{LANG06a,LANG09,PEDA10,ZECH17a}, allows to pattern large columnar defect arrays (CDAs). Briefly, thin Si stencil masks, perforated with square arrays of holes with diameters $D = 180 \pm 5$~nm, are mounted on top of the YBCO film. Direct contact between the mask and the surface of the bridge is avoided by a 1.5-$\mu$m-thick spacer. The arrays of about $670 \times 270$ holes with lattice constants $d = (302 \pm 2)$~nm (sample A) and about $400 \times 160$ holes with $d = (500 \pm 2)$~nm (sample B), respectively, cover the entire YBCO bridge. A possible misalignment of the mask parallel to the long sides of the prepatterned YBCO bridge is smaller than $0.3^\circ$ for the production of sample A and about $(1 \pm 0.5)^\circ$ in sample B. The irradiation is performed in a commercial ion implanter (High Voltage Engineering Europa B. V.) on a cooled sample stage with the ion beam applied parallel to the sample's $c$ axis and the dose is monitored by Faraday cups.

The electrical measurements are performed in a closed-cycle cryocooler mounted between the pole pieces of an electromagnet. A Cernox resistor \cite{HEIN98} together with a LakeShore 336 temperature controller is used for in-field temperature control to a stability of about 1 mK. The magnetic field, oriented perpendicular to the sample surface, is tuned by a programmable constant current source and measured with a calibrated Hall probe. Values are cross checked using a LakeShore 475 gaussmeter equipped with a HSE probe with a resolution of 0.1 $\mu$T, a zero offset $<10~\mu$T, and a reading accuracy $<0.1~\%$. The longitudinal current through the sample was generated by a Keithley 2400-LV constant-current source in both polarities to exclude thermoelectric signals and the longitudinal and transverse voltages measured simultaneously by the two channels of a Keithley 2182A nanovoltmeter.

To ensure a well-defined initial arrangement of vortices in the sample, the following protocol is applied for every datum of the measurements. First, the sample is slowly field cooled (FC) from the normal state down to the respective measurement temperature. Afterwards, all data are taken with multiple reversals of the polarity of the excitation current (30 times) and the magnetic field (four times) to improve the signal-to-noise ratio. The error bars in some of the figures mark the 95\% confidence intervals of the mean values of these multiple measurements. The magnetic field is applied instantaneously to overcome possible vortex pinning effects at the edges of the sample. The large demagnetization factor of our thin-film samples together with eddy currents induced by the rapid field switching leads to a full penetration of vortices into the sample. Some vortices are trapped in the CDs and others at interstitial positions in disordered Bose glass arrangement.

Finally, the sample was warmed up to 100~K and the FC procedure repeated to collect the next data set. Obviously, the ``virgin'' data collected right after FC might probe a different and more ordered vortex arrangement than those recorded after the rapid reversals of the field's polarity. However, we find the difference to be small and tentatively leaving out this first data set leads to very similar results.

\section{Results}

\begin{figure}[t]
\centering
\includegraphics*[width=\columnwidth]{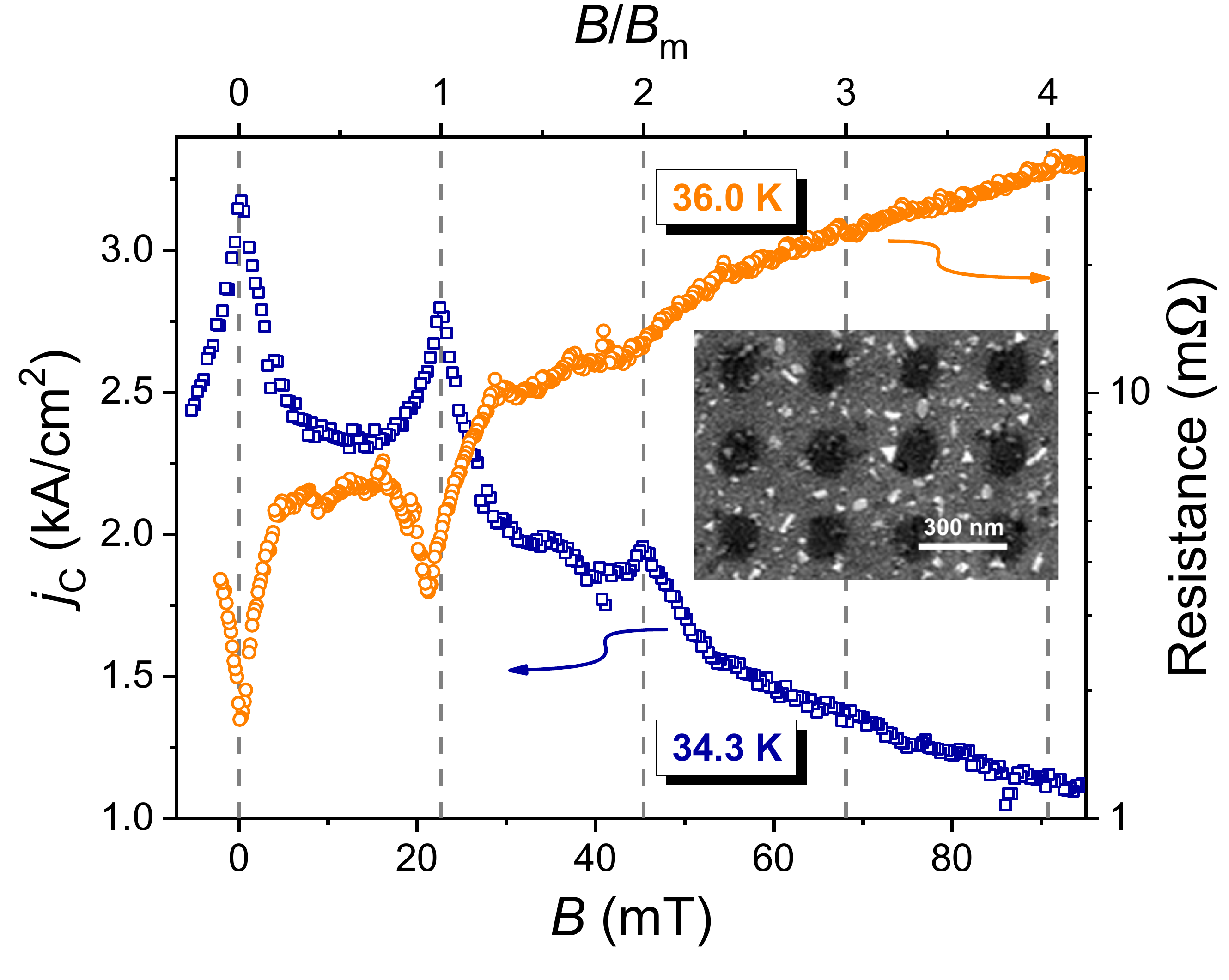}
\caption[]{Vortex commensurability effects in a YBCO thin film (sample A) with an artificial pinning landscape demonstrated by peaks in the critical current density and minima of the magnetoresistance. The top axis is normalized to the matching field $B_m = 22.7$~mT. To ensure equilibrium vortex arrangement the sample was heated to $T=100\ \mathrm{K}$ and then field cooled for every datum. The inset shows a scanning electron microscopy picture of the sample surface, where the black areas indicate the irradiated and, thus, non-superconducting defect columns.}
\label{fig:JcR}       % Give a unique label
\end{figure}

Periodic pinning structures in superconductors exhibit the so-called ``vortex matching'' effect, which can be detected in various physical quantities when the vortex lattice is commensurate with the lattice of pinning sites in the sample. For a square array of pinning sites the (first) matching field is given by
\begin{equation}
\label{eq:matching}
B_m = \frac{{{\phi _0}}}{{{d^2}}},
\end{equation}
where $\phi_0$ is the flux quantum and $d$ is the lattice constant of the square array.

As an example of commensurability effects in the longitudinal electrical transport properties, the critical current density $j_c$ and the resistance $R_{xx}$ of sample A ($d = 302$~nm) as a function of the magnetic field $B$ oriented perpendicular to the sample surface is shown in Fig.~\ref{fig:JcR}. To allow for a quasiequilibrium arrangement of vortices the data are collected with the respective magnetic field applied at 100~K before cooling the sample below $T_c$ for every datum \cite{HAAG14}. The distinct maxima in $j_c(B)$ and respective minima in $R_{xx}(B)$ are positioned at multiples $n$ of the matching field $n B_m = n (22.7 \pm 0.2)$~mT, according to Eq.~(\ref{eq:matching}) and taking the lattice parameter $d$ from measurements of the stencil mask in a scanning electron microscope (SEM). Interestingly, the positions of the peaks in $j_c(B)$ and the minima in $R_{xx}(B)$ coincide almost perfectly, despite that they are recorded at different temperatures and the former is static, whereas the latter is a dynamic probe of vortex commensurability effects. Similar demonstrations of vortex matching effects in thin YBCO films have been reported before \cite{CAST97,SWIE12,TRAS13,HAAG14,ZECH17a}.

\begin{figure}[t]
\centering
\includegraphics*[width=0.9\columnwidth]{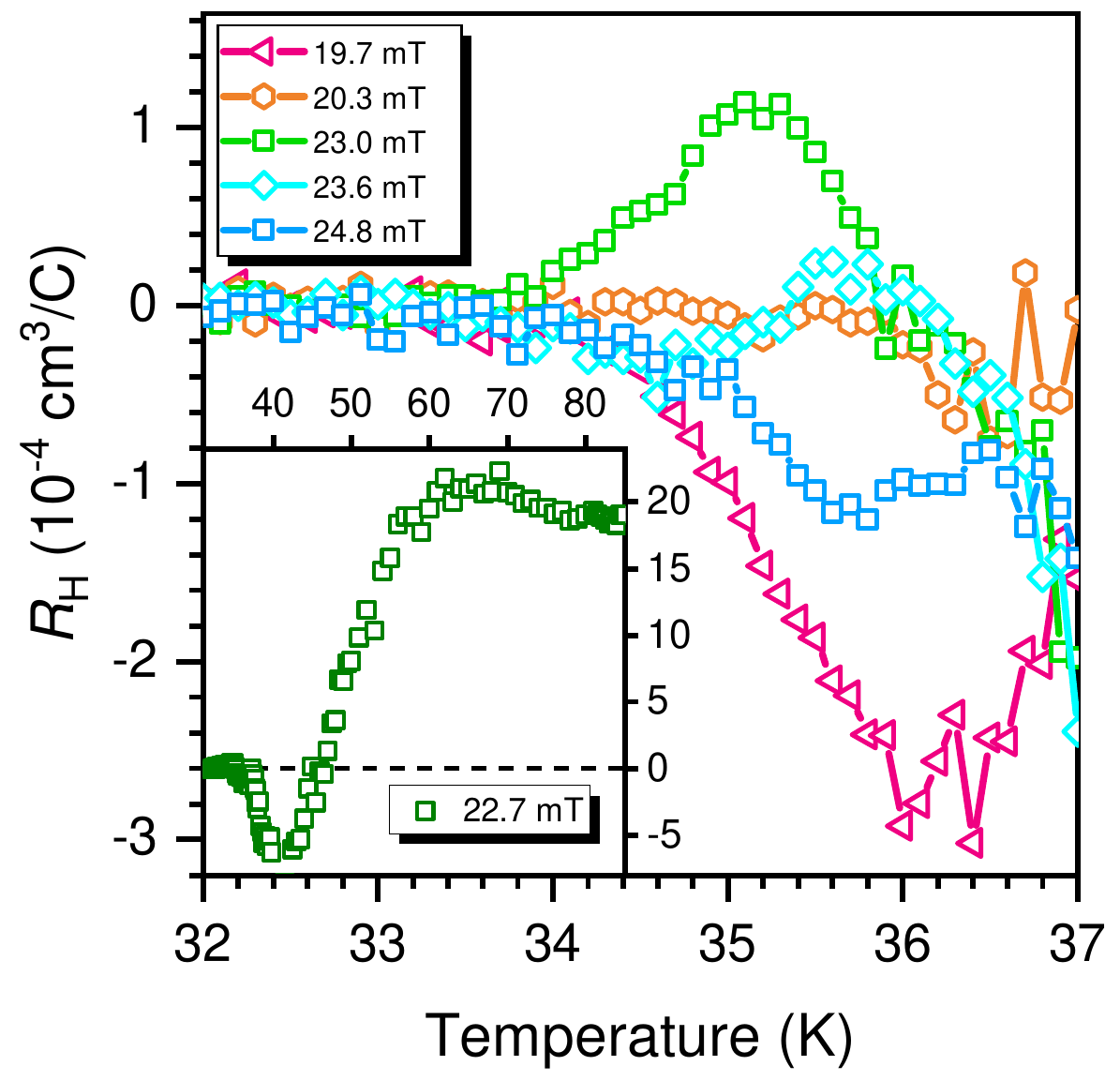}
\caption[]{Onset of the Hall coefficient of sample A at temperatures below the superconducting transition in various small magnetic fields. The matching field is $B_m = 22.7$~mT according to Eq.~(\ref{eq:matching}). Inset: Hall coefficient measured at $B_m$ over a wider temperature range.}
\label{fig:RHvsT}
\end{figure}

Correspondingly, sample B ($d = 500$~nm) shows a peak in $j_c$ at $B = (7.8 \pm 0.3)$~mT (obtained from a fit with background subtraction), where $B_m = 8.3$~mT, calculated from Eq.~(\ref{eq:matching}) using the mask's geometry as determined by SEM. The slight discrepancy is due to a broader and less pronounced matching peak because the intrinsic pinning of the YBCO films becomes more important at the wider CDA lattice and also to some uncertainty in the absolute values of the rather low magnetic fields.

The midpoint of the superconducting transition is lowered to $T_c \sim 47$~K ($T_c \sim 83$~K) in sample A (sample B) after irradiation, probably caused by some straggling of the ions after passing through the holes in the stencil mask and within the YBCO film itself. This leads to a, although minor, number of defects in the interstitial regions between the CDs and thus a reduction of $T_c$ due to the $d$-wave symmetry of the superconducting gap in YBCO. A similar observation was reported by other authors \cite{SWIE12}. Previous full-area irradiation experiments of YBCO films with He$^+$ ions \cite{SEFR01,LANG04} revealed that the carrier concentration remains unchanged and the irradiation does not lead to underdoping. The Hall effect remains positive in the normal state but the carrier mobility is reduced with increasing irradiation fluence \cite{LANG04}. Hence, the observation of high-mobility electron pockets in the Fermi surface of underdoped YBCO that provokes a sign change of the Hall coefficient at intense magnetic fields \cite{LEBO07} is not relevant for our present experiments.

The Hall effect in sample A around the superconducting transition in an applied magnetic field that corresponds to the matching field $B_m = 22.7$~mT is displayed in the inset of Fig.~\ref{fig:RHvsT}. At first sight, the temperature dependence of $R_H$ looks similar to what is found in pristine YBCO films \cite{LANG94}. The reversal of the Hall effect's sign in the vicinity of the normal to superconducting transition is generally observed at low and moderate magnetic fields and will be considered an intrinsic effect in this paper, although its explanation is still not consensual \cite{PUIC04}.

Closer inspection of the emergence of the Hall signal when pinning is overcome reveals a strikingly different behavior upon small variations of the magnetic field \cite{ZECH17}. Whereas close to the matching field, $R_H$ arises with positive sign, it comes up negative otherwise (see Fig.~\ref{fig:RHvsT}). At higher temperatures, the curves merge into the one shown in the inset of Fig.~\ref{fig:RHvsT}.

\begin{figure}[t]
\centering
\includegraphics*[width=\columnwidth]{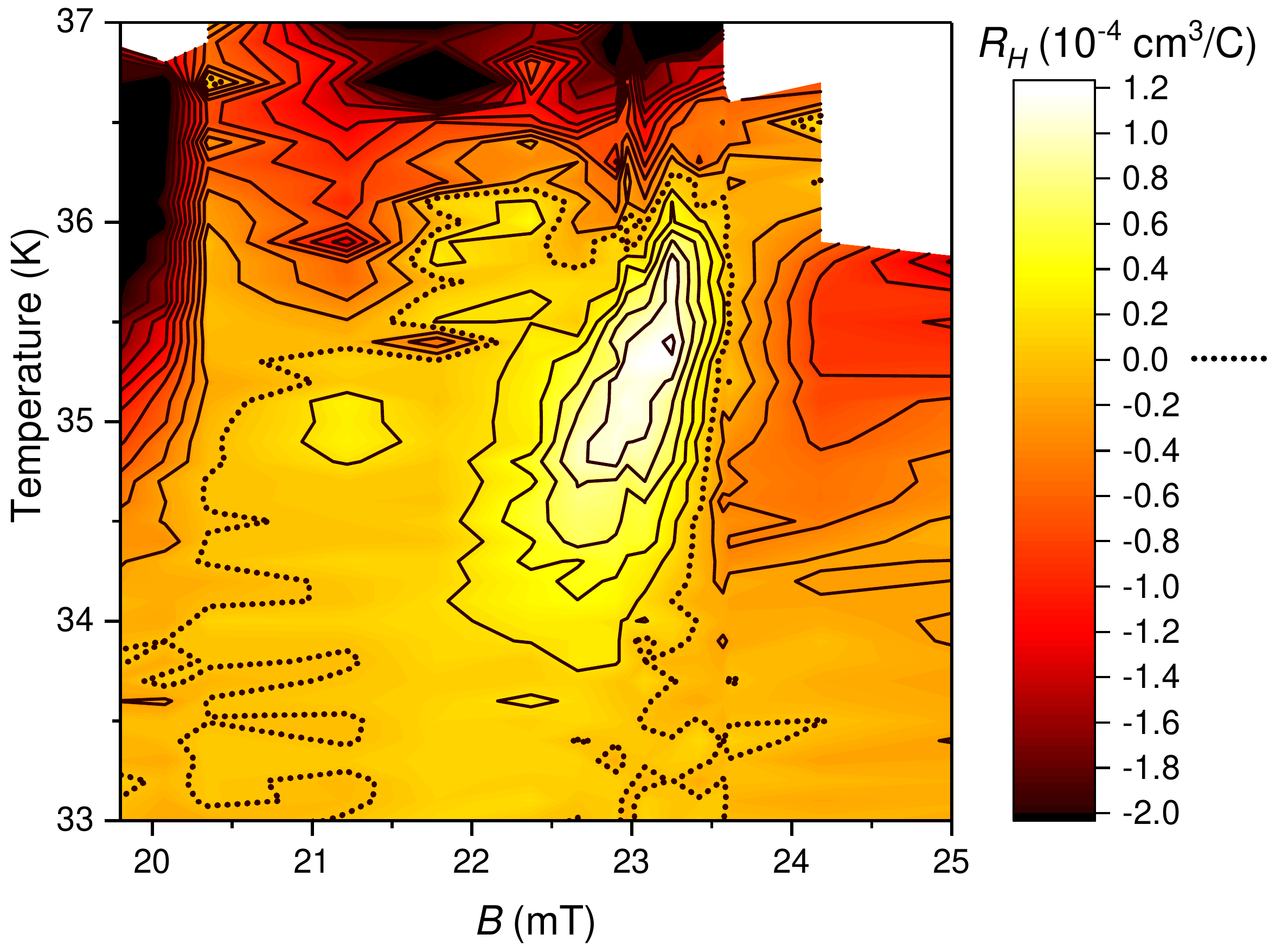}
\caption[]{Contour plot of the Hall coefficient at various temperatures and magnetic fields of a YBCO film with a CDA. The bright area corresponds to a maximum of the Hall effect at the matching field. The dotted line indicates the bifurcation, where the Hall effect emerges with different sign.}
\label{fig:contour}
\end{figure}

\begin{figure}[t]
\centering
\includegraphics*[width=0.9\columnwidth]{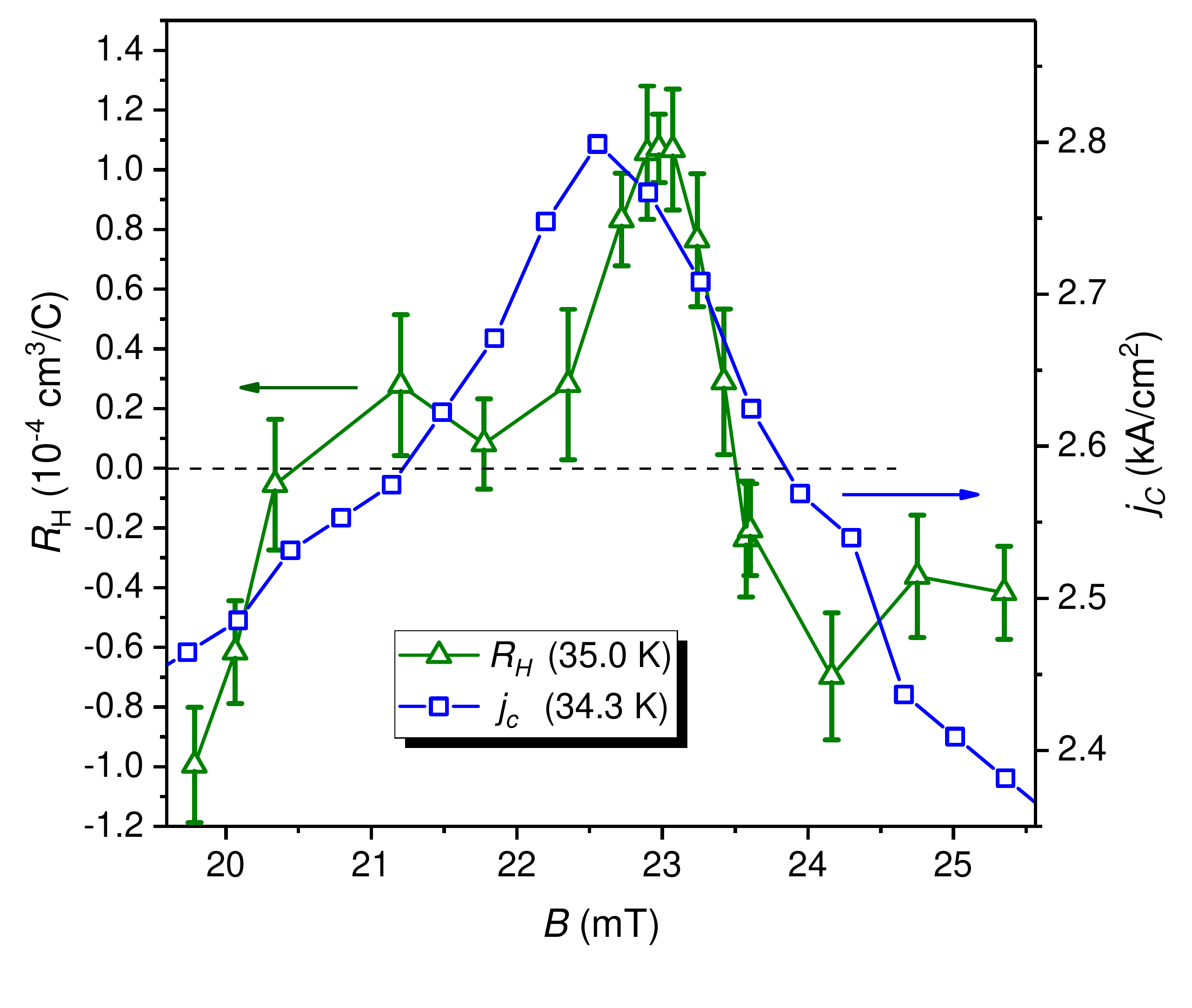}
\caption[]{Comparison of longitudinal and transverse vortex commensurability effects in the YBCO film with a CDA. Blue squares show the peak of the critical current at 34.3~K and green triangles show the magnitude of the Hall coefficient at 35.0~K near the matching field. Error bars indicate the 95~\% confidence intervals of the mean value.}
\label{fig:comp}       % Give a unique label
\end{figure}

Figure~\ref{fig:contour} provides a comprehensive overview of the Hall effect in various magnetic fields around $B_m$ and at different temperatures. Lighter colors represent a larger positive $R_H$, darker colors a larger modulus of negative $R_H$ and the dotted line marks the bifurcation between emergent Hall effects of different sign. Evidently, $R_H$ peaks around $B = 23\ \mathrm{mT} \sim B_m$ at temperatures $T = (35 \pm 1)$~K, corresponding to a reduced temperature range of $t = T/T_c = 0.74 \pm 0.02$. Outside the parameter range enveloped by the dotted line, $R_H$ is zero or negative up to $T \sim 47$~K, where it turns positive, characteristic for the normal state.

The peak in the Hall signal is compared to the well-established signature of vortex matching by a peak in $j_c(B)$ in Fig.~\ref{fig:comp} for sample A and in Fig.~\ref{fig:rh500} for sample B. By appropriate scaling it is possible to almost collapse the $R_H(B)$ and $j_c(B)$ curves onto each other in sample A, whereas the Hall peak is somewhat sharper in sample B. This indicates that commensurability effects in the Hall channel might be even more pronounced than for longitudinal transport.

To the best of our knowledge, a sign change of the Hall effect in a superconductor with a regular pinning array, limited to a narrow region around $B_m$, has not been observed before in other kinds of samples. However, additional sign changes of $R_H$ or the Hall conductivity $\sigma_{xy}=R_H B/(\rho_{xx}^2+R_H^2 B^2) \approx R_H B/\rho_{xx}^2$ have been reported in HgBa$_2$CaCu$_2$O$_6$ after heavy-ion irradiation \cite{KANG00} and in YBCO films in low magnetic fields due to pinning along twin boundaries \cite{GOB00}. Also, significant changes of $\sigma_{xy}$ due to pinning at twin boundaries have been revealed in YBCO single crystals \cite{DANN98}. These reports suggest that pinning can have an even stronger influence on the Hall channel than on the longitudinal transport, but they have considered disordered pinning sites only. The central result of our paper is the observation of a \emph{commensurate} Hall effect peak and sign change appearing at the matching field only.

\begin{figure}[t]
\centering
\includegraphics*[width=0.9\columnwidth]{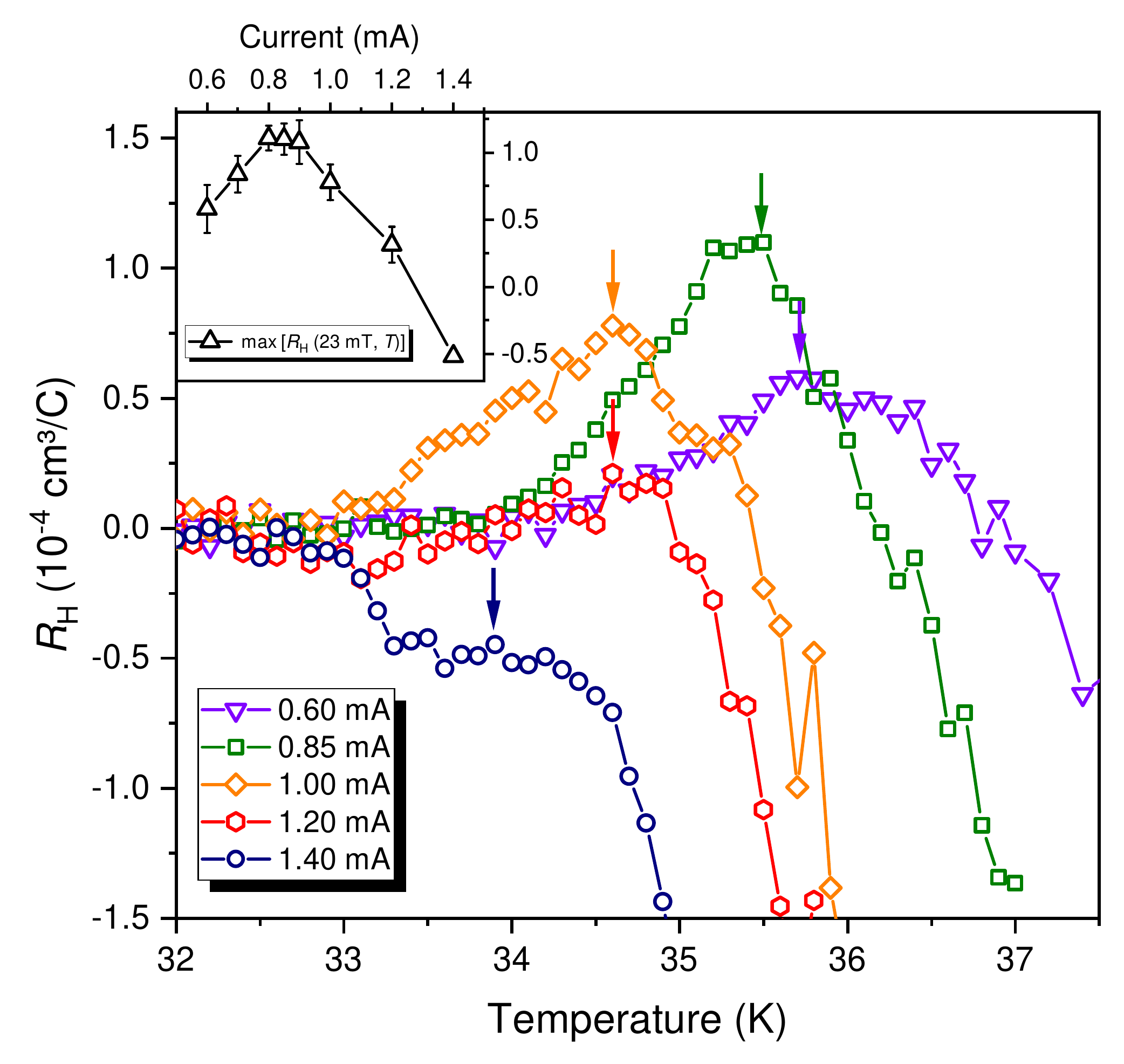}
\caption[]{Hall coefficient vs temperature at $B_m$, probed with different currents. Inset: Current dependence of the peak values of the $R_H(T)$ curves indicated by arrows in the main panel. Error bars indicate the 95~\% confidence intervals of the mean value.}
\label{fig:idep}
\end{figure}

Since the observed effect is rather small and, in general, Hall effect measurements in the mT range are delicate, a careful assessment of possible spurious effects is mandatory. Due to the large amount of data [16 independent $R_H(T)$ curves] represented in Fig.~\ref{fig:contour} and the effect's reproducibility in another sample with a different CDA an erratic effect can be ruled out. Typical 95~\% confidence intervals (assuming a normal distribution) calculated from data of the multiple reversals of current and magnetic field are displayed as error bars in Figs.~\ref{fig:comp}--\ref{fig:rh500} and confirm the relevance of our observations.

Nevertheless, a systematic error could arise from a slight misalignment of the transverse voltage probes, resulting in a transverse voltage drop $V_T$ originating from the longitudinal resistance, which is probed by the voltage $V_L$. Together with an asymmetry of the applied magnetic field $\Delta|B|=|B^+| - |B^-|$ this could mimic a Hall voltage. For sample A, $V_T/V_L < 0.014$ in the normal state above 100~K, $\Delta|B| < 90\,\mu$T, $V_L(+B_m) - V_L(-B_m) \approx 100$~nV, and thus a possible erroneous Hall voltage $V_H^{err} < 0.014 [V_L(+B_m) - V_L(-B_m)]/2 \approx 0.7$~nV---much smaller than $V_H(B_m, 35~\mathrm{K}) \approx 20$~nV.

Another effect that might be confused with our present observation of a Hall voltage is the occurrence of a transverse voltage due to guided vortex motion in patterned superconductors. For instance, in superconductors patterned with oblique microchannels for easy vortex flow, a substantial transverse voltage has been reported \cite{LAVI10} and confirmed that it vanishes when the easy channel is oriented parallel or perpendicular to the Lorentz force on the vortices \cite{DOBR16}. Similarly, square arrays of antidots \cite{SILH03,WORD04} or ferromagnetic dots \cite{VILL03a}, tilted with respect to the Lorentz force, allow for directional vortex channeling. Note that such guiding effects will result in an ``even'' Hall effect, $V_T(+B) = V_T(-B)$, which can be experimentally distinguished from the conventional (odd) Hall effect, albeit the combination of guiding and disordered pinning mechanisms can evoke an odd Hall voltage, too \cite{KOPN99,SHKL06}.

Vortex guiding effects were reported to result in minima of $V_T$ at multiples of $B_m$ due to enhanced commensurate pinning \cite{CHIA05}, in sharp contrast to the maxima of the Hall effect that coincide with maxima in pinning strength, reflected by the peaks in $j_c(B)$, as it is observed in our experiment (see Figs.~\ref{fig:comp} and \ref{fig:rh500}). Finally, in our experimental design particular care was exercised to avoid guiding effects by orienting the main axes of the CDA parallel to the current and to the Lorentz force, respectively. In addition, we could not trace an even Hall voltage in our measurements, which makes it unlikely that guiding of vortices does influence our observations.

Furthermore, vortex flow rectification effects are predicted in arrays of pins with asymmetric shape \cite{OLSO05,SAVE05} and in hexagonal lattices of symmetric pins with a spacial gradient \cite{REIC16}. Such effects can lead to a transverse voltage, which, however, would be canceled out by our measurement protocol that includes current and magnetic field reversal for every datum. Also, the necessary symmetry breaking is absent in our CDA.

The above considerations suggest that the commensurability peak and the sign change of $R_H(B)$ are related to the dynamic interaction of itinerant vortices with the pinning landscape of the CDA near the borderline where thermal activation can overcome the vortex pinning. To this end, an investigation of the nonlinearity of the observed effects by varying the current through the sample is illustrative and is shown in Fig.~\ref{fig:idep} for sample A and in the inset of Fig.~\ref{fig:rh500} for sample B. The value of the local maximum of $R_H(B_m,T)$ rapidly decreases with larger currents in both samples, pointing to the relevance of pinning, and eventually the Hall peak disappears at larger vortex velocities. Interestingly, a slight reduction of the peak can be seen also towards lower currents, but it has to be cautioned against experimental uncertainty, which increases under these low signals. Note that the reduction of the commensurability peak is in contrast to the intrinsic {\em negative} anomalous Hall effect, which becomes more prominent and extends to a wider temperature range in high currents \cite{LANG01,PUIC09}.

\section{Discussion}

The problem of vortex commensurability effects in the Hall channel has not been directly addressed so far. In conjunction with the anomalous negative Hall effect, observed in most HTSCs, it has been pointed out that the complex behavior of the Hall effect results from various additive contributions to the Hall conductivity $\sigma_{xy}=\sigma_{xy}^N+\sigma_{xy}^S+\sigma_{xy}^P$, where  $\sigma_{xy}^N$ represents a quasiparticle or vortex-core contribution, $\sigma_{xy}^S$ is a superconducting contribution, resulting from  hydrodynamic vortex effects and superconducting fluctuations, \cite{OTTE95,DORS92a,TROY93,KOPN93,NISH97} and $\sigma_{xy}^P$ allows for a pinning dependence of $\sigma_{xy}$. The sign of $\sigma_{xy}^N$ is the same as that of the normal-state Hall effect, i.e., positive in YBCO, but the sign of $\sigma_{xy}^S$ depends on details of the Fermi surface \cite{DORS92a,TROY93,KOPN93,OTTE95,NAGA98,BERG15}. The Hall effect's sign reversal and behavior in a wide range of magnetic fields in \emph{unpatterned} HTSCs can be quantitatively modeled \cite{PUIC04}.

The pinning contribution $\sigma_{xy}^P$ can evoke a second sign reversal of $R_H(T)$, provided that it has the opposite sign of $\sigma_{xy}^S$ and a similar magnitude. Kopnin and Vinokur  \cite{KOPN99} have proposed such a scenario in twinned YBCO films and Ikeda \cite{IKED99} has emphasized that the sign of  $\sigma_{xy}^P$ does depend on the dimensionality of the pinning, namely ${\rm sgn}(\sigma_{xy}^P)={\rm sgn}(\sigma_{xy}^S)$ for a vortex glass with point like disordered pinning sites and ${\rm sgn}(\sigma_{xy}^P) \ne {\rm sgn}(\sigma_{xy}^S)$ for a Bose glass, when disordered line like pinning centers dominate.

\begin{figure}[t]
\centering
\includegraphics*[width=\columnwidth]{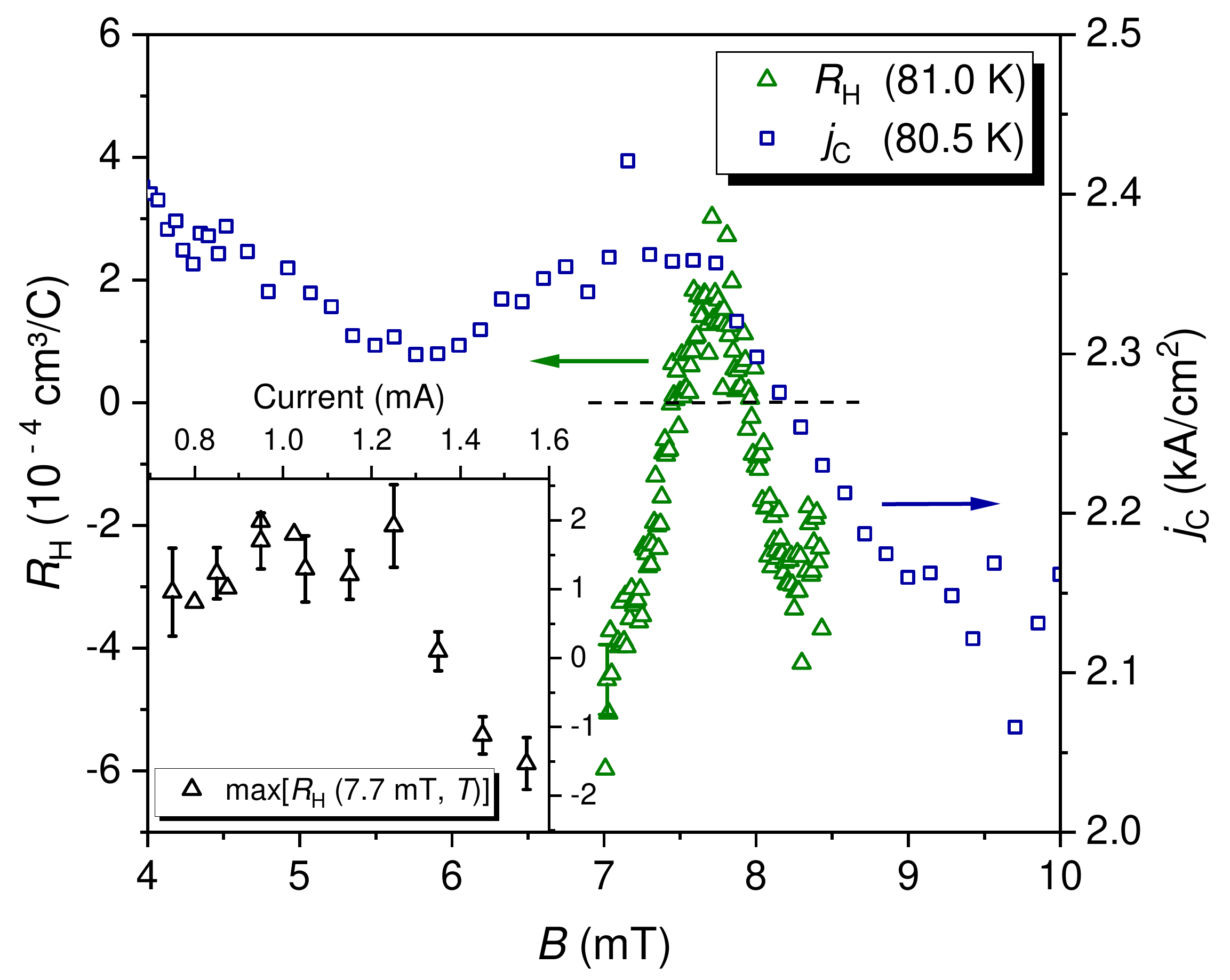}
\caption[]{Comparison of the peak of the critical current at 80.5~K (blue squares) and the Hall coefficient at 81.0~K (green triangles) in a sample with a wider-spaced CDA ($d = 500$~nm). Inset: The current dependence of the peak value of the $R_H(T)$ curves, measured at 7.7~mT, the maximum of $R_H(B)$. Error bars indicate the 95~\% confidence intervals of the mean value.}
\label{fig:rh500}
\end{figure}

\begin{figure}[t]
\centering
\includegraphics*[width=\columnwidth]{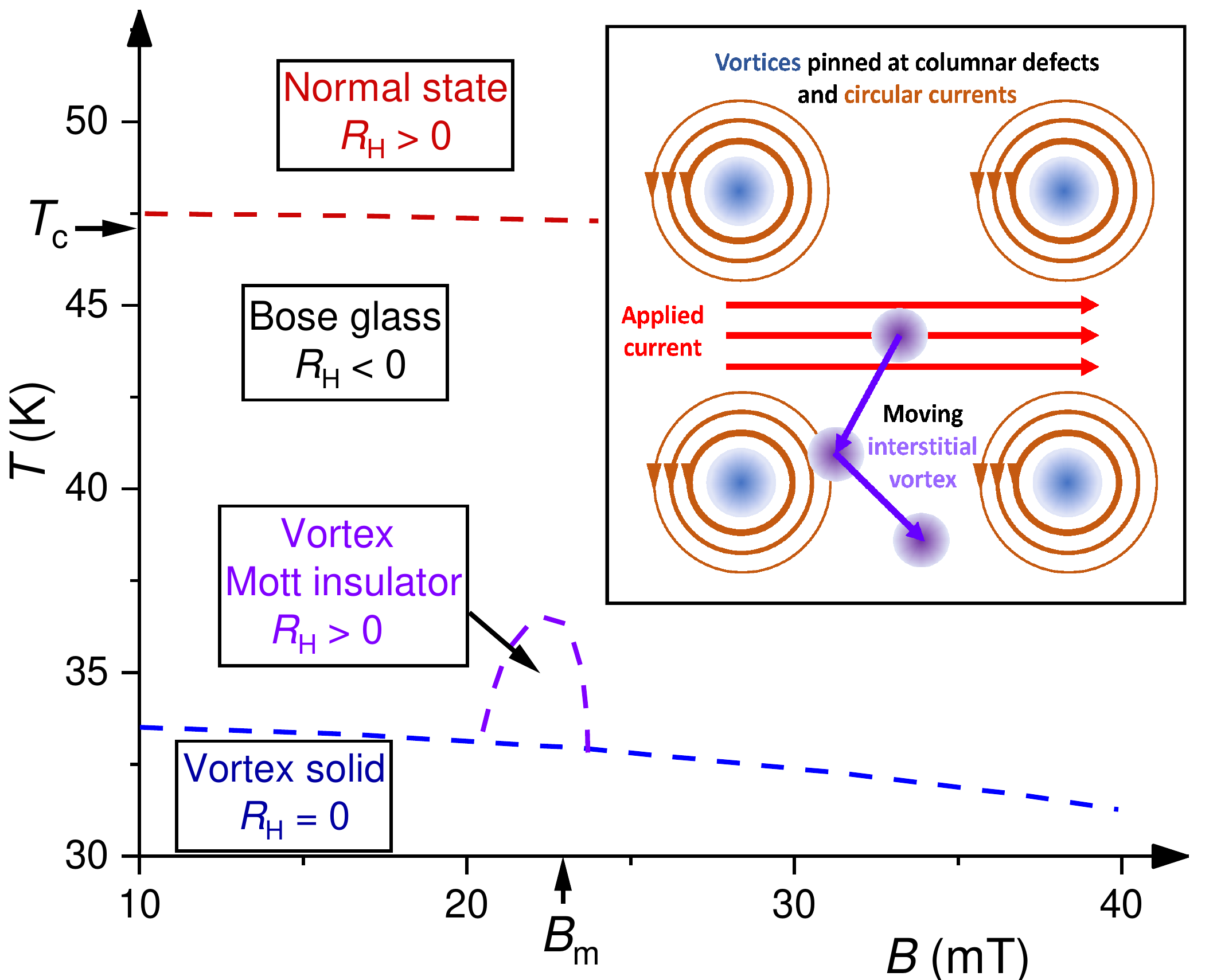}
\caption[]{Schematic diagram of the Hall coefficient's behavior at various temperatures and magnetic fields, based on data in sample A. Broken lines indicate a vanishing Hall effect. The inset shows a sketch of the possible situation in the vortex Mott insulator state, when a few mobile interstitial vortices are deflected by the circular currents around the pinned vortices in the CDA. Adapted from Nakai \etal \cite{NAKA11}.}
\label{fig:diagram}
\end{figure}

In our samples the situation is related but somewhat different from the above-mentioned models. Our observations are summarized in a schematic phase diagram of the Hall effect in Fig.~\ref{fig:diagram}. Lowering the temperature across $T_c$ leads to the aforementioned sign change of $R_H$ from its positive normal-state value to a negative one in a Bose glass phase. Note that this sign change roughly coincides with the midpoint $T_c$ at zero field. At lower temperatures, the system might undergo a transition to a vortex Mott insulator state that is confined to a region around the matching field. There, the majority of vortices are trapped in the CDA, but nevertheless a few interstitial vortices are present, dominating the transport properties, and evoking the peak of $R_H > 0$ in the \emph{B-T} plane displayed in Fig.~\ref{fig:contour}. This regime is determined by several boundary conditions. Along the $B$ axis, vortex trapping is maximized at the commensurability field $B_m$. In the temperature regime, the vortex Hall effect approaches zero when all vortices are strongly pinned into a vortex solid and, at higher temperatures, the vortex Mott insulator state is gradually destroyed by thermal fluctuations leading to an increasing number of interstitial vortices with an intrinsically negative $R_H$. Hence, the positive peak of the Hall effect is limited to a rather narrow temperature interval.

A possible scenario for the sign change of $R_H$ in a vortex Mott insulator is sketched in the inset of Fig.~\ref{fig:diagram}. Nakai \etal \cite{NAKA11} have proposed that moving vortices are deflected from their intrinsic trajectories by the circular currents surrounding pinned vortices. Their simulations revealed a sign change of the Hall voltage for disordered pinned vortex arrangements. The effect might be even enhanced in a regular array of pinned vortices---a situation that arises at the matching field in our sample. Note that the negative Hall effect implies an upstream component of vortex motion with respect to the applied current. This approach is somewhat related to the concept of the Magnus force acting on moving vortices due to an interaction of circular shielding currents with the laminar transport current \cite{AO93,SONI97,ZHU97}. When the commensurable vortex arrangement is destroyed by either an off-matching external field or in high current densities, the number of mobile interstitial vortices is larger and their intrinsic Hall behavior prevails. Tuning the system between such different states of vortex matter might thus change the delicate vortex dynamics in the Hall channel and lead to the observed sign change of the Hall effect at $B_m$.

Alternative scenarios of deviations of vortex trajectories from their preferred direction parallel to the Lorentz force have been considered, too. Giamarchi and Le Doussal \cite{GIAM96} have proposed a transverse critical current for a moving vortex glass, and simulations by Reichhardt and co-workers \cite{REIC98,REIC00} have identified a transverse critical force in a smectic vortex phase existing in superconductors with regular pinning arrays and related transverse commensurability effects with the number of moving rows of vortices between neighboring rows of pinning sites \cite{REIC08}. Kolton \etal \cite{KOLT99} have predicted a Hall noise in driven vortex lattices, which disappears at high driving forces, in some agreement with our results presented in Figs.~\ref{fig:idep} and \ref{fig:rh500}. Although these theoretical approaches predict  transverse displacements of individual vortices or vortex channels, they are of arbitrary nature and would average out at the time scales of our transport measurements.

To provide the necessary symmetry breaking in the problem of moving vortex ensembles that can produce a finite Hall voltage, a Hall term has to be considered in the equation of motion of a vortex, as it is for instance incorporated in the Bardeen-Stephen model \cite{BARD65}. Considering the Magnus force is another option, but little is known about the Magnus force in a vortex system with a regular pinning array. Along these lines, our observation of a commensurability peak in the Hall effect in superconductors with a pin lattice might spark additional theoretical efforts to develop a more detailed picture of vortex motion in regular pin arrays by including additional force terms.

\section{Conclusions}

A regular array of defect columns in thin YBCO films not only gives rise to maxima in the critical current and minima in the resistance, but also to a manifestation of commensurability in the transverse Hall signal. A peak of the Hall coefficient and a related sign change from its intrinsic negative values in the mixed state to positive appear in a narrow magnetic field range around the matching field. Previous theoretical results have predicted a sign change of the Hall signal due to strong pinning of a vortex Bose glass but no predictions have been made for the vortex Hall effect in a superconductor with a regular pinning array and for the Hall effect near a vortex Mott metal/insulator transition. We have confirmed that our findings are \emph{not} related to guided vortex motion and instead suggest that the transverse Hall voltage can be a subtle probe for vortex dynamics in a periodic pinning landscape that needs further attention.\\

\section*{Acknowledgments}

We appreciate the help of K. Haselgr\"ubler with the ion implanter. Illuminating discussions with V.~M.~Vinokur, A.~Silhanek, R. W\"ordenweber, and O.~V.~Dobrovolskiy are acknowledged. This article is based upon work from COST Action CA16218 (NANOCOHYBRI), supported by COST (European Cooperation in Science and Technology). M.D. acknowledges the European Erasmus Mundus (Target II) program for financial support.

\end{document}